\documentclass[%
 twocolumn,
 aps,pra,
 amsmath,amssymb,
 superscriptaddress,
 nofootinbib,
 reprint,%
]{revtex4}
\usepackage{stmaryrd}
\usepackage{color}
\usepackage{array}
\usepackage{enumitem}
\usepackage{mathpazo}
\usepackage{setspace}
\usepackage{bm}
\usepackage{amssymb}
\usepackage{amsthm}
\usepackage{mathtools}
\usepackage{multirow}
\usepackage{cases}
\usepackage[colorlinks=true,linkcolor=blue,citecolor=blue,pdfauthor={ },pdftitle={ },pdfsubject={ },pdfkeywords={ }]{hyperref}
\usepackage{pgfplots}
\usepackage{lipsum}
\usepackage{youngtab}
\usepackage{subfigure}
\usepackage{verbatim}
\usepackage[capitalise]{cleveref}

\usepackage{algorithm}
\usepackage{algpseudocode}

\newtheorem{definition}{Definition}
\newtheorem{proposition}[definition]{Proposition}
\newtheorem{lemma}[definition]{Lemma}

\newtheorem{theorem}[definition]{Theorem}

\newtheorem{corollary}[definition]{Corollary}
\newtheorem{conjecture}[definition]{Conjecture}

\newtheorem{remark}[definition]{Remark}
\newtheorem{example}[definition]{Example}
\newtheorem{question}[definition]{Question}

\def\Dbar{\leavevmode\lower.6ex\hbox to 0pt
{\hskip-.23ex\accent"16\hss}D}
\makeatletter
\def\url@leostyle{%
  \@ifundefined{selectfont}{\def\UrlFont{\sf}}{\def\UrlFont{\small\ttfamily}}}
\makeatother
\DeclareMathOperator{\tr}{tr} %
\DeclareMathOperator{\Tr}{Tr} %
\DeclareMathOperator{\rank}{rank}

\def\bcj{\begin{conjecture}}
\def\ecj{\end{conjecture}}
\def\bcr{\begin{corollary}}
\def\ecr{\end{corollary}}
\def\bd{\begin{definition}}
\def\ed{\end{definition}}
\def\bea{\begin{eqnarray}}
\def\eea{\end{eqnarray}}
\def\bem{\begin{enumerate}}
\def\eem{\end{enumerate}}
\def\bex{\begin{example}}
\def\eex{\end{example}}
\def\bim{\begin{itemize}}
\def\eim{\end{itemize}}
\def\bl{\begin{lemma}}
\def\el{\end{lemma}}
\def\bpf{\begin{proof}}
\def\epf{\end{proof}}
\def\bpp{\begin{proposition}}
\def\epp{\end{proposition}}
\def\bqu{\begin{question}}
\def\equ{\end{question}}
\def\br{\begin{remark}}
\def\er{\end{remark}}
\def\bt{\begin{theorem}}
\def\et{\end{theorem}}

\def\btb{\begin{tabular}}
\def\etb{\end{tabular}}

\newcommand{\nc}{\newcommand}

 \nc{\bA}{{\bf A}} \nc{\bB}{{\bf B}} \nc{\bC}{{\bf C}}
 \nc{\bD}{{\bf D}} \nc{\bE}{{\bf E}} \nc{\bF}{{\bf F}}
 \nc{\bG}{{\bf G}} \nc{\bH}{{\bf H}} \nc{\bI}{{\bf I}}
 \nc{\bJ}{{\bf J}} \nc{\bK}{{\bf K}} \nc{\bL}{{\bf L}}
 \nc{\bM}{{\bf M}} \nc{\bN}{{\bf N}} \nc{\bO}{{\bf O}}
 \nc{\bP}{{\bf P}} \nc{\bQ}{{\bf Q}} \nc{\bR}{{\bf R}}
 \nc{\bS}{{\bf S}} \nc{\bT}{{\bf T}} \nc{\bU}{{\bf U}}
 \nc{\bV}{{\bf V}} \nc{\bW}{{\bf W}} \nc{\bX}{{\bf X}}
 \nc{\bZ}{{\bf Z}}

\nc{\cA}{{\cal A}} \nc{\cB}{{\cal B}} \nc{\cC}{{\cal C}}
\nc{\cD}{{\cal D}} \nc{\cE}{{\cal E}} \nc{\cF}{{\cal F}}
\nc{\cG}{{\cal G}} \nc{\cH}{{\cal H}} \nc{\cI}{{\cal I}}
\nc{\cJ}{{\cal J}} \nc{\cK}{{\cal K}} \nc{\cL}{{\cal L}}
\nc{\cM}{{\cal M}} \nc{\cN}{{\cal N}} \nc{\cO}{{\cal O}}
\nc{\cP}{{\cal P}} \nc{\cQ}{{\cal Q}} \nc{\cR}{{\cal R}}
\nc{\cS}{{\cal S}} \nc{\cT}{{\cal T}} \nc{\cU}{{\cal U}}
\nc{\cV}{{\cal V}} \nc{\cW}{{\cal W}} \nc{\cX}{{\cal X}}
\nc{\cZ}{{\cal Z}}

\nc{\hA}{{\hat{A}}} \nc{\hB}{{\hat{B}}} \nc{\hC}{{\hat{C}}}
\nc{\hD}{{\hat{D}}} \nc{\hE}{{\hat{E}}} \nc{\hF}{{\hat{F}}}
\nc{\hG}{{\hat{G}}} \nc{\hH}{{\hat{H}}} \nc{\hI}{{\hat{I}}}
\nc{\hJ}{{\hat{J}}} \nc{\hK}{{\hat{K}}} \nc{\hL}{{\hat{L}}}
\nc{\hM}{{\hat{M}}} \nc{\hN}{{\hat{N}}} \nc{\hO}{{\hat{O}}}
\nc{\hP}{{\hat{P}}} \nc{\hR}{{\hat{R}}} \nc{\hS}{{\hat{S}}}
\nc{\hT}{{\hat{T}}} \nc{\hU}{{\hat{U}}} \nc{\hV}{{\hat{V}}}
\nc{\hW}{{\hat{W}}} \nc{\hX}{{\hat{X}}} \nc{\hZ}{{\hat{Z}}}

\newcommand{\bra}[1]{\langle#1|}
\newcommand{\ket}[1]{|#1\rangle}

\def\Dbar{\leavevmode\lower.6ex\hbox to 0pt
{\hskip-.23ex\accent"16\hss}D}
\begin{document}

\def\be{\begin{eqnarray}}
\def\ee{\end{eqnarray}}

\newcommand{\ca}{\mathcal A}

\newcommand{\cb}{\mathcal B}
\newcommand{\cc}{\mathcal C}
\newcommand{\cd}{\mathcal D}
\newcommand{\ce}{\mathcal E}
\newcommand{\cf}{\mathcal F}
\newcommand{\cg}{\mathcal G}
\newcommand{\ch}{\mathcal H}
\newcommand{\ci}{\mathcal I}
\newcommand{\cj}{\mathcal J}
\newcommand{\ck}{\mathcal K}
\newcommand{\cl}{\mathcal L}
\newcommand{\cm}{\mathcal M}
\newcommand{\cn}{\mathcal N}
\newcommand{\co}{\mathcal O}
\newcommand{\cp}{\mathcal P}
\newcommand{\cq}{\mathcal Q}
\newcommand{\calr}{\mathcal R}
\newcommand{\cs}{\mathcal S}
\newcommand{\ct}{\mathcal T}
\newcommand{\cu}{\mathcal U}
\newcommand{\cv}{\mathcal V}
\newcommand{\cw}{\mathcal W}
\newcommand{\cx}{\mathcal X}
\newcommand{\cy}{\mathcal Y}
\newcommand{\cz}{\mathcal Z}

\newcommand{\sa}{\mathscr{A}}
\newcommand{\sm}{\mathscr{M}}

\newcommand{\fa}{\mathfrak{a}}  \newcommand{\Fa}{\mathfrak{A}}
\newcommand{\fb}{\mathfrak{b}}  \newcommand{\Fb}{\mathfrak{B}}
\newcommand{\fc}{\mathfrak{c}}  \newcommand{\Fc}{\mathfrak{C}}
\newcommand{\fd}{\mathfrak{d}}  \newcommand{\Fd}{\mathfrak{D}}
\newcommand{\fe}{\mathfrak{e}}  \newcommand{\Fe}{\mathfrak{E}}
\newcommand{\ff}{\mathfrak{f}}  \newcommand{\Ff}{\mathfrak{F}}
\newcommand{\fg}{\mathfrak{g}}  \newcommand{\Fg}{\mathfrak{G}}
\newcommand{\fh}{\mathfrak{h}}  \newcommand{\Fh}{\mathfrak{H}}
\newcommand{\fraki}{\mathfrak{i}}       \newcommand{\Fraki}{\mathfrak{I}}
\newcommand{\fj}{\mathfrak{j}}  \newcommand{\Fj}{\mathfrak{J}}
\newcommand{\fk}{\mathfrak{k}}  \newcommand{\Fk}{\mathfrak{K}}
\newcommand{\fl}{\mathfrak{l}}  \newcommand{\Fl}{\mathfrak{L}}
\newcommand{\fm}{\mathfrak{m}}  \newcommand{\Fm}{\mathfrak{M}}
\newcommand{\fn}{\mathfrak{n}}  \newcommand{\Fn}{\mathfrak{N}}
\newcommand{\fo}{\mathfrak{o}}  \newcommand{\Fo}{\mathfrak{O}}
\newcommand{\fp}{\mathfrak{p}}  \newcommand{\Fp}{\mathfrak{P}}
\newcommand{\fq}{\mathfrak{q}}  \newcommand{\Fq}{\mathfrak{Q}}
\newcommand{\fr}{\mathfrak{r}}  \newcommand{\Fr}{\mathfrak{R}}
\newcommand{\fs}{\mathfrak{s}}  \newcommand{\Fs}{\mathfrak{S}}
\newcommand{\ft}{\mathfrak{t}}  \newcommand{\Ft}{\mathfrak{T}}
\newcommand{\fu}{\mathfrak{u}}  \newcommand{\Fu}{\mathfrak{U}}
\newcommand{\fv}{\mathfrak{v}}  \newcommand{\Fv}{\mathfrak{V}}
\newcommand{\fw}{\mathfrak{w}}  \newcommand{\Fw}{\mathfrak{W}}
\newcommand{\fx}{\mathfrak{x}}  \newcommand{\Fx}{\mathfrak{X}}
\newcommand{\fy}{\mathfrak{y}}  \newcommand{\Fy}{\mathfrak{Y}}
\newcommand{\fz}{\mathfrak{z}}  \newcommand{\Fz}{\mathfrak{Z}}

\newcommand{\cfg}{\dot \fg}
\newcommand{\cFg}{\dot \Fg}
\newcommand{\ccg}{\dot \cg}
\newcommand{\circj}{\dot {\mathbf J}}
\newcommand{\circs}{\circledS}
\newcommand{\jmot}{\mathbf J^{-1}}

\newcommand{\rmd}{\mathrm d}
\newcommand{\mca}{\ ^-\!\!\ca}
\newcommand{\pca}{\ ^+\!\!\ca}
\newcommand{\peq}{^\Psi\!\!\!\!\!=}
\newcommand{\lt}{\left}
\newcommand{\rt}{\right}
\newcommand{\HN}{\hat{H}(N)}
\newcommand{\HM}{\hat{H}(M)}
\newcommand{\Hv}{\hat{H}_v}
\newcommand{\cyl}{\mathbf{Cyl}}
\newcommand{\lag}{\left\langle}
\newcommand{\rag}{\right\rangle}
\newcommand{\Ad}{\mathrm{Ad}}
\newcommand{\trace}{\mathrm{tr}}
\newcommand{\bbc}{\mathbb{C}}
\newcommand{\AC}{\overline{\mathcal{A}}^{\mathbb{C}}}
\newcommand{\Ar}{\mathbf{Ar}}
\newcommand{\uc}{\mathrm{U(1)}^3}
\newcommand{\M}{\hat{\mathbf{M}}}
\newcommand{\spin}{\text{Spin(4)}}
\newcommand{\id}{\mathrm{id}}
\newcommand{\Pol}{\mathrm{Pol}}
\newcommand{\Fun}{\mathrm{Fun}}
\newcommand{\bp}{p}
\newcommand{\act}{\rhd}
\newcommand{\data}{\lt(j_{ab},A,\bar{A},\xi_{ab},z_{ab}\rt)}
\newcommand{\datao}{\lt(j^{(0)}_{ab},A^{(0)},\bar{A}^{(0)},\xi_{ab}^{(0)},z_{ab}^{(0)}\rt)}
\newcommand{\deltadata}{\lt(j'_{ab}, A',\bar{A}',\xi_{ab}',z_{ab}'\rt)}
\newcommand{\background}{\lt(j_{ab}^{(0)},g_a^{(0)},\xi_{ab}^{(0)},z_{ab}^{(0)}\rt)}
\newcommand{\sgn}{\mathrm{sgn}}
\newcommand{\vth}{\vartheta}
\newcommand{\rmi}{\mathrm{i}}
\newcommand{\bfmu}{\pmb{\mu}}
\newcommand{\bfnu}{\pmb{\nu}}
\newcommand{\bfm}{\mathbf{m}}
\newcommand{\bfn}{\mathbf{n}}
\newcommand{\perk}{\mathfrak{S}_k}
\newcommand{\dens}{\mathrm{D}}
\newcommand{\iden}{\mathbb{I}}
\newcommand{\End}{\mathrm{End}}

\newcommand{\sz}{\mathscr{Z}}
\newcommand{\sk}{\mathscr{K}}

\title{Maximum entropy methods for quantum state compatibility problems}

\author{Shi-Yao Hou}%
\affiliation{College of Physics and Electronic Engineering, Center for Computational Sciences,  Sichuan Normal University, Chengdu 610068, China}
\affiliation{Shenzhen SpinQ Technology Co., Ltd., Shenzhen, China}

\author{Zipeng Wu}
\affiliation{Department of Physics, The Hong Kong University of Science and Technology, Clear Water Bay, Kowloon, Hong Kong, China}

\author{Jinfeng Zeng}
\affiliation{Department of Physics, The Hong Kong University of Science and Technology, Clear Water Bay, Kowloon, Hong Kong, China}
\affiliation{Beijing Academy of Quantum Information Sciences,  Beijing 100193,  China}

\author{Ningping Cao}
\affiliation{Institute for Quantum Computing, University of Waterloo, Waterloo, Ontario, Canada}

\author{Chenfeng Cao}
\affiliation{Department of Physics, The Hong Kong University of Science and Technology, Clear Water Bay, Kowloon, Hong Kong, China}

\author{Youning Li}
\email[]{liyouning@cau.edu.cn}
\affiliation{College of Science, China Agricultural University, Beijing, 100080, People's Republic of China}

\author{Bei Zeng}
\email[]{zengb@ust.hk}
\affiliation{Department of Physics, The Hong Kong University of Science and Technology, Clear Water Bay, Kowloon, Hong Kong, China}

\date{\today}

\begin{abstract}
Inferring a quantum system from incomplete information is a common problem in many aspects of quantum information science and applications, where the principle of maximum entropy (MaxEnt) plays an important role. The quantum state compatibility problem asks whether there exists a density matrix $\rho$ compatible with some given measurement results. Such a compatibility problem can be naturally formulated as a semidefinite programming (SDP), which searches directly for the existence of a $\rho$. However, for large system dimensions, it is hard to represent $\rho$ directly, since it needs too many parameters.  In this work, we apply MaxEnt to solve various quantum state compatibility problems, including the quantum marginal problem. An immediate advantage of the MaxEnt method is that it only needs to represent $\rho$ via a relatively small number of parameters, which is exactly the number of the operators measured. Furthermore, in case of incompatible measurement results, our method will further return a witness that is a supporting hyperplane of the compatible set. Our method has a clear geometric meaning and can be computed effectively with hybrid quantum-classical algorithms.
\end{abstract}

\maketitle

\section{Introduction}

The connection of statistical mechanics and information theory encodes a fundamental logic of science.
In standard statistical mechanics\cite{pathria2016statistical}, it has been generally accepted that canonical distribution of equilibrium corresponds to the state with maximum entropy, after early work of Boltzman\cite{boltzmann1877beziehung}.
With the foundation and boom of information theory in 1940s\cite{shannon1948mathematical}, Jaynes developed the connotation of entropy in statistical mechanics and endow new information-theoretical meaning of this concept\cite{jaynes1957information,jaynes1957information2}.
Lying at the heart is the principle of maximum entropy (MaxEnt), saying that the probability distribution that best represents the current state of knowledge is the one with maximum entropy.
In information theory, MaxEnt is excellent in dealing with the inference problem of incomplete information, which is ubiquitous in science.
With linear constraints, the general solution for MaxEnt adopts an exponential form, which is widely studied in terms of the information geometry methods~\cite{csiszar2004information}.
MaxEnt finds a wide range of applications in areas such as data classification~\cite{palmieri2013objective}, density estimation~\cite{botev2011generalized}, natural language processing~\cite{ratnaparkhi1998maximum}, and parameter estimation in artificial intelligence~\cite{malouf2002comparison}.

With the emergence of quantum information theory in 1970s\cite{holevo1973bounds}, MaxEnt developed its own new meaning and applications in this new field, by extending the information entropy to von Neumann entropy of quantum states.
A natural problem in quantum information theory is the quantum state inference from local information. Inference with complete information is the so-called quantum tomography~\cite{buvzek1998reconstruction}, while the incomplete information version is quantum maximum entropy inference (qMaxEnt).
qMaxEnt is applied to a broad range of problem in quantum information, such as entanglement detection\cite{vrehavcek2001iterative}.
Later studies on quantum information geometry further connected the qMaxEnt to correlations in many-body quantum systems\cite{zhou2008irreducible,niekamp2013computing}.

In this paper we focus on the use of qMaxEnt to solve the quantum state compatibility problem. For a quantum system of $n$ qubits, consider a set of Hermitian operators $\{A_1,A_2,\cdots,A_m\}$, and corresponding a set of real numbers $\{a_1,a_2,\cdots,a_m\}$. The quantum state compatibility problem asks whether there exists a quantum state $\rho$ of the system, such that $a_i=\Tr(\rho A_i)$. This problem can be naturally formulated as a semidefinite programming (SDP), to find the solutions to the following linear matrix inequalities (LMIs)
\begin{align}
&\quad \forall A_i \quad\Tr (\rho A_i)=a_i, \\
& \quad \rho\succeq 0.
\label{eq:sdp}
\end{align}
If the solution to the LMIs exist, that is all $a_i$s are compatible, we will obtain a density matrix $\rho$ satisfying $\Tr (\rho A_i)=a_i$. Otherwise the quantum state compatibility problem has no solution. When $n$ is large, however, it is hard to represent $\rho\succeq 0$ which has exponential number of parameters in $n$.

In practice, normally $m$ is only polynomial in $n$ (e.g. $A_i$s are local operators), we can instead to use the qMaxEnt method. That is, we can consider an operator $H$ of the form
\begin{align}
H=\sum c_i A_i,
\end{align}
and state of the form
\begin{align}
\rho'=\frac{e^{-H}}{\Tr(e^{-H})},
\end{align}
then by MaxEnt, a compatible state $\rho$ exists if and only if $\rho'$ exists, which satisfies $a_i=\Tr(\rho' A_i)$. An immediate advantage of this qMaxEnt compared to the SDP algorithm is that $\rho'$ only needs $m$ parameters to represent.

\begin{figure}
\includegraphics[scale=0.3]{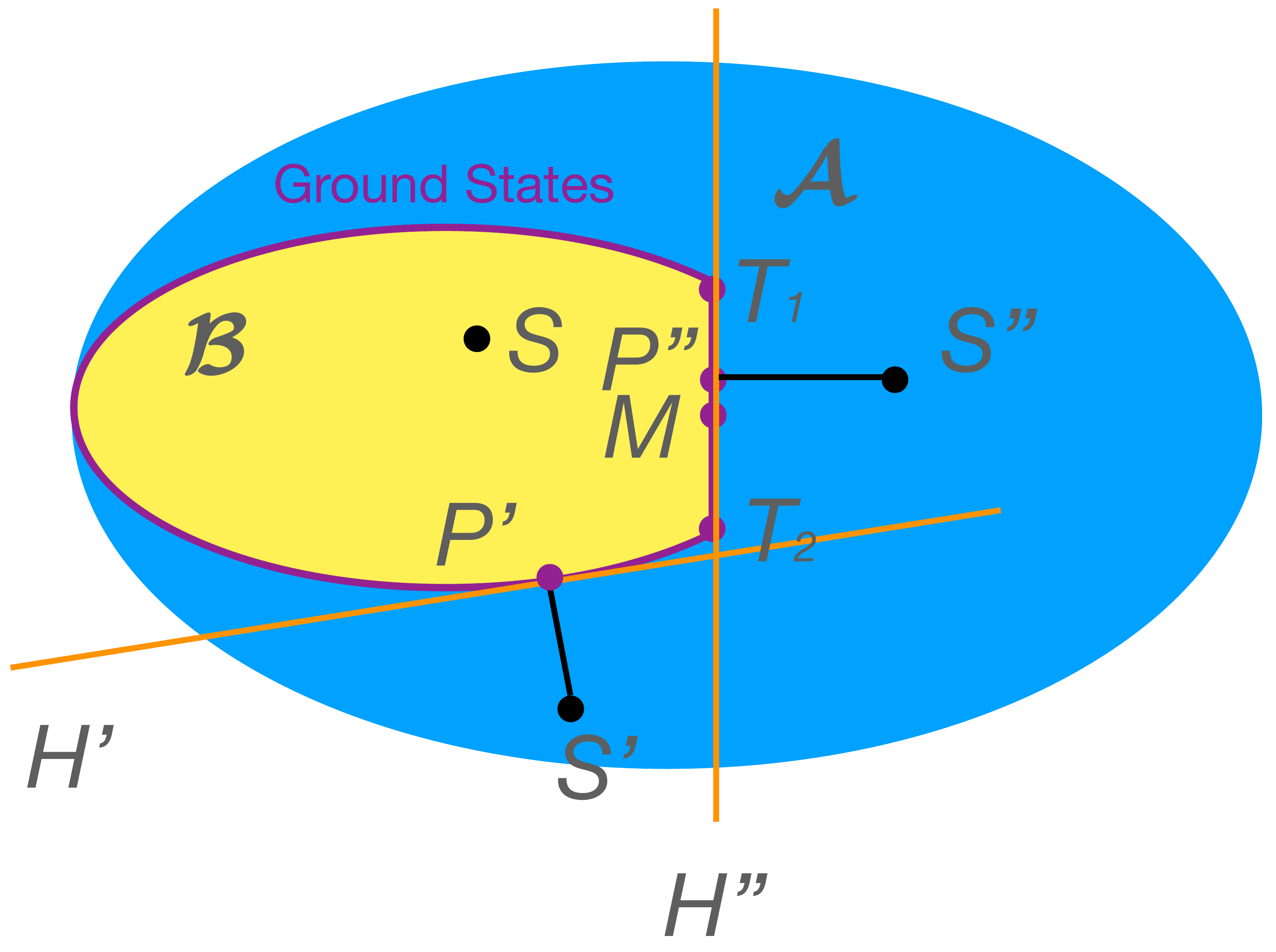}
\caption{\textbf{Geometric interpretation of the maximum entropy method.} $\mathcal{A}$ denotes the set of all $a_i$s. $\mathcal{B}$ denotes the set of all $a_i$s corresponding to some compatible state $\rho$, which is a convex set. The boundary of $\mathcal{B}$ represents the ground states of some Hamiltonian $H=\sum c_i A_i$. Flat part of the boundary of $\mathcal{B}$ (e.g., $T_1$-$T_2$) corresponding to degenerate ground states of some $H$. For $a_i$s in $\mathcal{B}$ (e.g., point $S$), the qMaxEnt method finds compatible states of the form $\rho'=\frac{e^{-H}}{\Tr(e^{-H})}$. If the corresponding quantum state compatibility problem has no solution, then $a_i$s must be outside of $\mathcal{B}$ (e.g., points $S'$, $S''$). In this case, the qMaxEnt algorithm returns some density matrices corresponding to the boundary points of $\mathcal{B}$ (e.g., points $P'$, $P''$). $P'$ corresponds to the unique ground state of some Hamiltonian $H'$, which is the witness detecting the incompatibility of  $S'$ (the supporting hyperplane is tangent to $\mathcal{B}$ at point $P'$). $P''$ corresponds to a mixture state in the degenerate ground state space of some $H''$ since it falls to the flat region on the boundary of $\mathcal{B}$. In this case, a further analysis of the spectrum of the corresponding density matrix of $P''$ will lead to the witness $H''$, which is the supporting hyperplane of $\mathcal{B}$ that intersects $P''$.}
\label{fig:geoint}
\end{figure}

However, if the algorithm cannot return any $\rho'$, then it could due to the effectiveness and the convergence of numerical algorithms used. In other words, if we can find $\rho'$ satisfying $a_i=\Tr(\rho' A_i)$ to certain precision, we can then say that the answer to the compatibility problem is ``yes" and $\rho'$ is a compatible state. However, if we fail to find $\rho'$ satisfying $a_i=\Tr(\rho' A_i)$ to certain precision, we will need to further justify that the answer to the compatibility problem is indeed ``no". To do so, we will further find a \textit{witness}. The witness distinguishes the given incompatible $a_i$s and all the compatible $a_i$s.  The idea of our method is illustrated in \cref{fig:geoint}. Notice that when the SDP failed, it only returns ``no", without any further information.

The most difficult part of our method is to calculate $\tr(A_i\frac{e^{-H}}{\Tr(e^{-H})})$ for some $H$, where one can use some machine learning algorithms. In recent years, machine learning was introduced in to help with quantum information processing\cite{cao2020supervised, 10.1117/1.AP.4.2.026004}.
Quantum machine learning concepts, which is directly related to qMaxEnt, has been developed, such as quantum Boltzmann machine (QBM)\cite{amin2018quantum}, especially for the case when there is no hidden layer. Furthermore,
with the incoming of the so-called noisy-intermedia-scale quantum (NISQ) era~\cite{preskill2018quantum}, many hybrid quantum-classical algorithms (such as the variational QBMs\cite{zoufal2021variational,shingu2021boltzmann}, the variational quantum code searcher\cite{cao2022quantumvariational}), which may be more efficient on near-term quantum devices and be able to replace the subroutines of traditional algorithm, has been developed. Of particular relevance are the hybrid quantum-classical algorithms that prepare the thermal state of the form $\frac{e^{-H}}{\Tr(e^{-H})}$~\cite{zeng2021VQHD, verdon2019quantum, wu2019variational, wang2021variational, Chowdhury2020VQTS, zoufal2021variational}.  All these algorithms can then help to scale our method with the existence of near-term quantum computers.

We organize our paper as follows: in Sec.~\ref{sec:algo}, we introduce the details of our new algorithms that solve the quantum state compatibility problems based on qMaxEnt; in Sec.~\ref{sec:res}, we apply  our algorithms to solve some quantum state compatibility problems for random measurement operators; in Sec.~\ref{sec:qmp}, we further apply our method to solve a special kind of the quantum state compatibility problem, i.e. the quantum marginal problem; in Sec.~\ref{sec:var}, we discuss the hybrid quantum-classical algorithms, which can help to scale our method; finally we conclude in Sec.~\ref{sec:con}.

\section{The maximum entropy method}\label{sec:algo}

For a quantum system of $n$ qubits, consider a set of Hermitian operators $\{A_1,A_2,\cdots,A_m\}$, and a corresponding set of real numbers $\{a_1,a_2,\cdots,a_m\}$. The quantum state compatibility problem asks whether there exists a quantum state $\rho$ of the system, such that $a_i=\Tr(\rho A_i), \forall i$.
The MaxEnt states that there exists a state $\rho$, which satisfies
\begin{equation}
a_i=\Tr (\rho A_i),
\end{equation}
if and only if there exists a maximum entropy state $\rho'$ such that
\begin{equation}
\Tr(\rho' A_i)=a_i.
\end{equation}
The maximum entropy state has the form
\begin{equation}
\rho'=\frac{e^ {-H}}{\Tr(e^{-H)}},
\label{eq:dm}
\end{equation}
where
\begin{equation}\label{eq:ham}
H=\sum_i c_i A_i.
\end{equation}

In this setting, $\rho'$ is determined by vecter $\vec{c}=\{c_1,c_2,\cdots,c_m\}$.
Therefore, if we can find the maximum entropy state, we can conclude that there is a solution to the compatibility problem. Otherwise, the compatible $\rho$ does not exist. Now the compatibility problem is transformed into finding the $c_i$s.

\subsection{The loss function}

We can then set up a loss function:
\begin{equation}
f(\vec{c})=\sum_i \left[a_i-\Tr \rho'(\vec{c} )A_i\right]^2.
\label{eq:loss}
\end{equation}
Obviously, if there does exist a $\rho'$, such that $\Tr(\rho' A_i)=a_i, \forall i$, the loss function $f$ will reach its minimum 0.Therefore, the problem becomes an optimization problem. Nowadays, optimization problems are usually solved by using gradient-based algorithms.
The basic idea of these algorithms is to define an objective function and find its minimum. Roughly, the gradient-based optimization algorithms can be described as follows.  For a function $f(\vec{c})$, where $\vec{c}=(c_1,c_2,\cdots,c_m)$ is a vector of parameters. In order to find the minimum of such function, we can randomly choose an initial value $\vec{c}^{0}$ and then iteratively update it to optimum. For the $k$-th iteration, update $c^{k}$ to $c^{k+1}$ by
\begin{equation}
\vec{c}^{k+1}=\vec{c}^k-\alpha\nabla f(\vec{c}),
\end{equation}
where $\alpha \in \mathbb{R}$ is a scalar. If $\alpha$ is small enough, it could be guaranteed that $f(\vec{c}^{k+1})\leq f(\vec{c}^k)$. Usually, $\alpha$ is determined by line search algorithms. To approach better performance and convergence, algorithms such as Broyden-Fletcher-Goldfarb-Shanno (BFGS) algorithm \cite{bfgs_b,bfgs_f,bfgs_g,bfgs_s} or conjugate gradient \cite{hestenes1952methods} are used more frequently.

One alternative loss function is based on the quantum Boltzmann machine \cite{amin2018quantum}. Quantum Boltzmann machine is a new machine learning approach proposed based on Boltzmann distribution of a quantum Hamiltonian. The basic idea of quantum Boltzmann machine is that for a quantum system described by Hamiltonian $H$ with adjustable parameters $H(\vec{\theta})$, of which the general form is decided by the system, by measuring certain operator $\Lambda_{\mathbf{v}}$, we can obtain the Boltzmann probability distribution $P_{\mathbf{v}}$:
\begin{equation}
P_{\mathbf{v}}=\frac{\tr (\Lambda_{\mathbf{v}} e^{-H})}{\tr e^{-H}}.
\end{equation}
The loss function is
\begin{equation}
\mathcal{L}=-\sum_{\mathbf{v}} P_{\mathbf{v}}^{\text {data }} \log \frac{\tr\left[\Lambda_{\mathbf{v}} e^{-H}\right]}{\tr\left[e^{-H}\right]}=-\sum_{\mathbf{v}}P_{\mathbf{v}}^{\text {data }}\log P_{\mathbf{v}},
\end{equation}
where $P_{\mathbf{v}}^{\text {data}}$ is the probability distribution of training data. This loss function is the cross entropy loss. If the two probability distribution are exactly the same, the loss function reaches its minimum.

For our problem, we can assume $P^{\text{data}}_{\mathbf{v}}$ be the measured result $a_i=\Tr (\rho A_i)$, and $P_{\mathbf{v}}$ be $\Tr(\rho' A_i)$, then the loss function is
\begin{equation}
\mathcal{L}=-a_i \log \Tr(\rho' A_i),
\label{eq:qbm0}
\end{equation}
However, $P^{\text{data}}_{\mathbf{v}}$ and $P_{\mathbf{v}}$ in loss function are probability distributions, that is, for each $\mathbf{v}$,
\begin{align}
0\leq &P_{\mathbf{v}}\leq 1,\\
0\leq &P^{\text{data}}_{\mathbf{v}}\leq 1,
\end{align}
and
\begin{align}
\sum_\mathbf{v}P_{\mathbf{v}}&=1,\\
\sum_\mathbf{v}P^{\text{data}}_{\mathbf{v}}&=1.
\end{align}
Obviously, if $A_i$s are chosen to be product of Pauli Matrices, the expectation value for any density operator $\rho$ is
\begin{equation}
-1\leq 1\Tr(\tilde{\rho}_m A_i)\leq 1.
\end{equation}
Therefore, \cref{eq:qbm0} can not be optimized (note that for logarithm function, the variable $\Tr(\rho A_i)$ must be positive). Therefore, we have to make $a_i$ and $\Tr(\rho A_i)$ probability distributions, that is, apart from the condition that $A_i$s should be complete,
\begin{equation}
\sum_i A_i= I,
\label{eq:id}
\end{equation}
and for any $\rho$
\begin{equation}
\tr (\rho A_i)\geq 0.
\label{eq:semidef}
\end{equation}
To satisfy \cref{eq:id} and \cref{eq:semidef}, we just need to make $A_i$s being a collection of positive semi-definite operators that sums to identity. A set of such operators is also called positive operator valued measures (POVMs). The condition of $A_i$s being complete means the POVMs should be chosen information complete (IC). More specifically, we can choose  $A_i$s to be symmetric information complete POVMs (SIC-POVMs).
One way of constructing SIC-POVMs could be found in Ref. \cite{SALAZAR2012325}. With $A_i$s being chosen to be SIC-POVMs, both
$a_i$ and $\Tr(\rho A_i)$ become a probability distribution. The optimization methods could be used to optimize the loss function \cref{eq:qbm0}.

Besides the the gradient descent, which changes the parameters $\vec{c}^i$ as a whole in every iteration, there are other ways of finding the minimum value of a function. One algorithm for solving such a problem is proposed in \cite{niekamp2013computing} which is based on alternating optimization. The basic idea of alternating optimization is to optimize one parameter each time.

\subsection{The witness}

However, one problem with gradient-based algorithms is that they often fall into local minima. For our problem, if the maximum entropy state is found, the problem is solved. But, if the maximum entropy state is not found, we can not arbitrarily say that the maximum entropy state does not exist. Although according to our tests, it is highly probable that we can find the maximum entropy state if it exists. Here, we give a geometric interpretation of this problem.

Let $\mathcal{A}$ be the set of all possible $a_i$s, $\mathcal{B}$ be the set for all $a_i$s with compatible $\rho$. The shape of $\mathcal{B}$ is very complicated, but one thing known is that the set is complex. The boundary of $\mathcal{B}$ are ground states of certain Hamiltonian with the form $H=\sum_i x_i A_i$, as shown in \cref{fig:geoint}. When the optimization is performed, the maximum entropy state is obviously limited in space $\mathcal{B}$. Therefore, if there exists a global state with compatible $\rho$, e.g., point $S$, then we can find a solution. If the given $\rho$ is not reduced density matrices of a certain global density matrix, a point $P$ which minimize the function $f(\vec{x})$ will be found. Since the definition of function $f(\vec{x})$ is actually the Euclidean distance of optimum point $P$ and the target point $S'$, $P$ is actually the point nearest to $S'$, which means $P$ must be on the boundary of $\mathcal{B}$. Therefore, $P$ represents a ground state (if $H$ is not degenerate) or a mixture of degenerate ground states (if $H$ is degenerate).  As mentioned before, there is no guarantee that the optimization will approach the point $S$ or $P$. But, since $P$ is a ground state or a mixture of the ground states of the Hamiltonian $H$, we can form a witness. Through point $P'$, we can draw a tangent hyper plane (in a two dimension space shown in \cref{fig:geoint}, the hyper plane is a line) with function
\begin{equation}
\frac{x-a_1}{c_1}+\frac{y-a_2}{c_2}+\cdots=0
\end{equation}

One thing has to be noticed here is the degeneracy. The boundary of $\mathcal{B}$ is composed of all ground states. For certain cases, the Hamiltonian $H$ could be degenerate. As shown in \cref{fig:geoint}. If the point of the maximum entropy state is point $P'$, $H'$ serves as the witness. However, if the point of all reduced density matrices is the point $S''$, then the nearest point is $P''$ will be found. The Hamiltonian corresponding to $P''$ is degenerate, that is, there are more than two states has the same lowest energy. By checking the rank of the corresponding density matrix or just diagonalizing the density matrix, we can decide whether the Hamiltonian is degenerate. As shown in \cref{fig:geoint}, all the states on the boundary between the points $T_1$ and $T_2$ are degenerated, and each is a mixture of different ground states. Those pure ground states are the states corresponding to $T_1$ and $T_2$. Suppose the pure states corresponding to $T_1$ and $T_2$ are $|\psi_{T_1}\rangle$ and $|\psi_{T_2}\rangle$, respectively, then we can generate a new density matrix:
\begin{equation}
\rho_M=\frac{1}{2}(|\psi_{T_1}\rangle\langle\psi_{T_1}|+|\psi_{T_2}\rangle\langle\psi_{T_2}|),
\end{equation}
which corresponds to the point $M$, in the middle of $T_1$ and $T_2$. Using the same optimization method to find a Hamiltonian $H''$, for which
\begin{equation}
\rho_M=\frac{e^{-H''}}{\Tr e^{-H''}}.
\end{equation}
Here, the witness is the hyperplane (line in 2-dimension space) through all $T_i$s, as shown in \cref{fig:geoint}.

We now summarize our algorithm in Algorithm 1 based on the loss function given in Eq.~\eqref{eq:loss}.

\begin{algorithm*}[!htb]
  \begin{algorithmic}
    \caption{The Maximum Entropy Method}
      \Require Hermitian operator $A_i$s and expectation $a_i$.
      \Ensure A maximum entropy state $\rho' =\frac{e^{-H}}{\Tr e^{-H}}$ satisfies $a_i=\Tr \rho' A_i$
        \State  Randomly initialize coefficients $\vec{c}=\{c_i\}$.
        \State $\rho \gets \frac{e^{-H(\vec{c})}}{\Tr(e^{-H(\vec{c})})}$.
        \State Compute the loss function $f(\vec{c})=\sum_{i}(a_i-\Tr \rho A_i)^2$.

        \While{$f(\vec{c})$ has not converged}
        		\State Estimate the gradient $\nabla f(\vec{c}) $ through finite differencing.
		\State $\vec{c} = \vec{c} - \alpha \nabla f(\vec{c})$.
		\State $\rho \gets \frac{e^{-H(\vec{c})}}{\Tr(e^{-H(\vec{c})})}$.
		\State Compute the loss function $f(\vec{c})$.
        \EndWhile
        \State $\rho' \gets \rho$.
        \State Return $\vec{c}$, $f(\vec{c})$, $\rho'$.

        \If {$f(\vec{c}) <=\epsilon $}
         \State  desired $\rho'$ is found
         \ElsIf {$\rank(\rho')=1$}
         \State Witness is the hyperplane:
         \begin{equation}
\frac{x-a_1}{c_1}+\frac{y-a_2}{c_2}+\cdots=0
\end{equation}
         \Else
         \State Find all pure states of $\rho'$:$|\psi_1\rangle,|\psi_2\rangle,\cdots,|\psi_n\rangle$
         \State Generate an equally mixed state
         \begin{equation}
         \rho"=\frac{1}{n}\sum_{i=1}^n |\psi_i\rangle\langle\psi_i|,
         \end{equation}
         \State Find $H"$ and $\vec{c"}$ using the above method, so that
         \begin{equation}
         \rho"=\frac{e^{-H"}}{\Tr e^{-H"}},
         \end{equation}
         \State Witness is the hyperplane:
         \begin{equation}
			\frac{x-a_1}{c"_1}+\frac{y-a_2}{c"_2}+\cdots=0
		\end{equation}

         \EndIf

  \end{algorithmic}
\end{algorithm*}

\section{Results}
\label{sec:res}

We now apply our algorithm to study some concrete quantum state compatibility problems. We start from a single case where $m=2$, i.e. only two measurement operators $A_1, A_2$. In this case, we are looking at the set of $\{a_1, a_2\}$. The compatible set is then given by the points $\{\tr(\rho A_1), \tr(\rho A_2)\}$, which is known to be the same as the joint numerical range of $A_1,A_2$ \cite{bonsall_duncan_1971,bonsall1973numerical,jonckheere1998differential,gutkin2004convexity,gutkin2013joint}, given by
\begin{equation}
W(A_1,A_2)=\{(\langle x|A_1|x\rangle,\langle x|A_2|x\rangle)| |x\rangle \in \mathcal{H}, \langle x|x\rangle=1\}.
\end{equation}

In this case we can represent the Hamiltonian $H$ as
\begin{equation}
H=\cos\theta A_1+\sin \theta A_2.
\end{equation}
The ground state of $H$ is $|g\rangle$, then  the trajectory of
\begin{equation}
(\langle g|A_1|g\rangle, \langle g|A_2|g\rangle)
\end{equation}
as $\theta$ goes from 0 to $2\pi$ forms the boundary of $W(A_1,A_2)$.

\begin{figure*}[htb]
\includegraphics[scale=0.3]{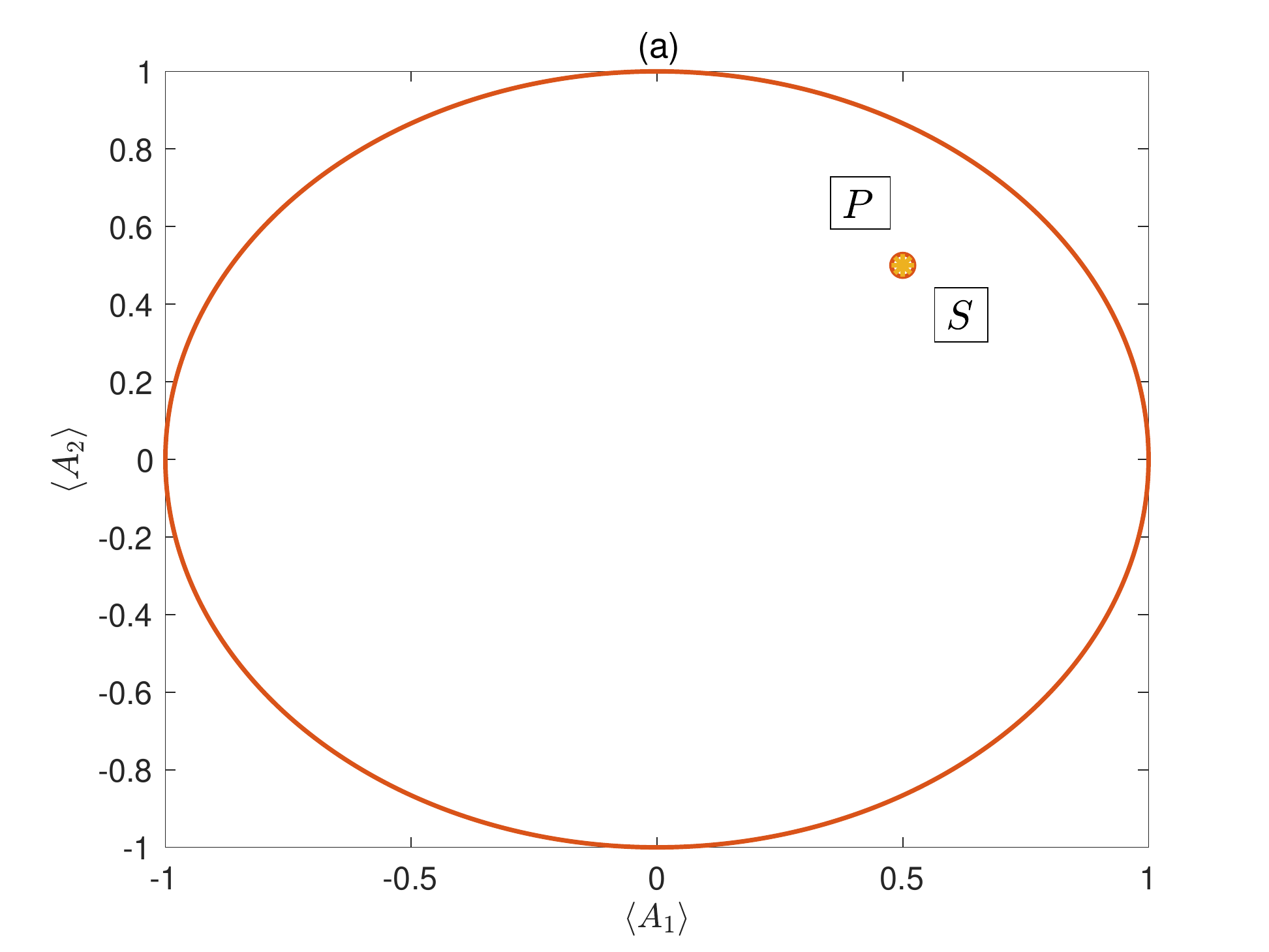}\includegraphics[scale=0.3]{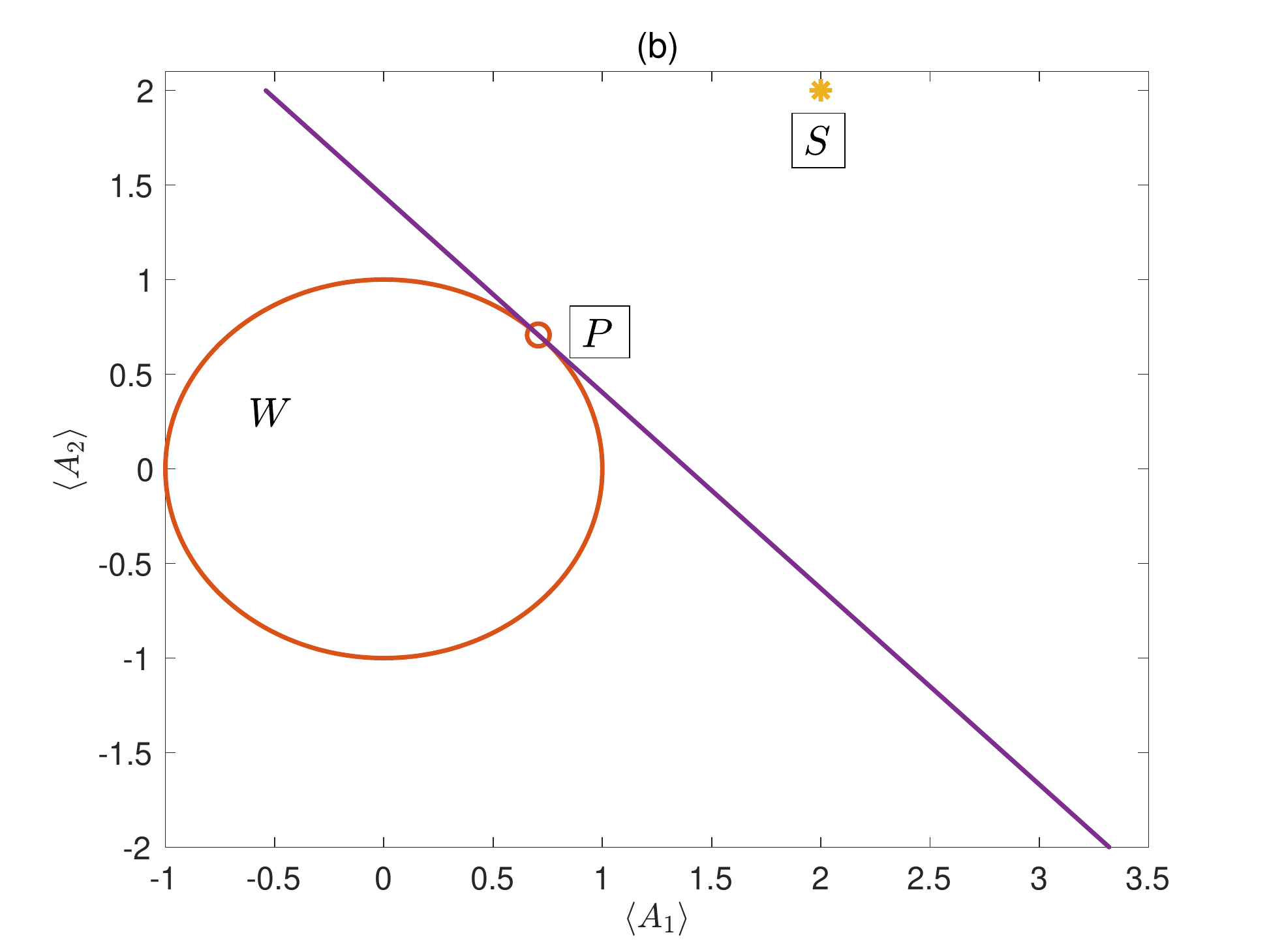}\includegraphics[scale=0.3]{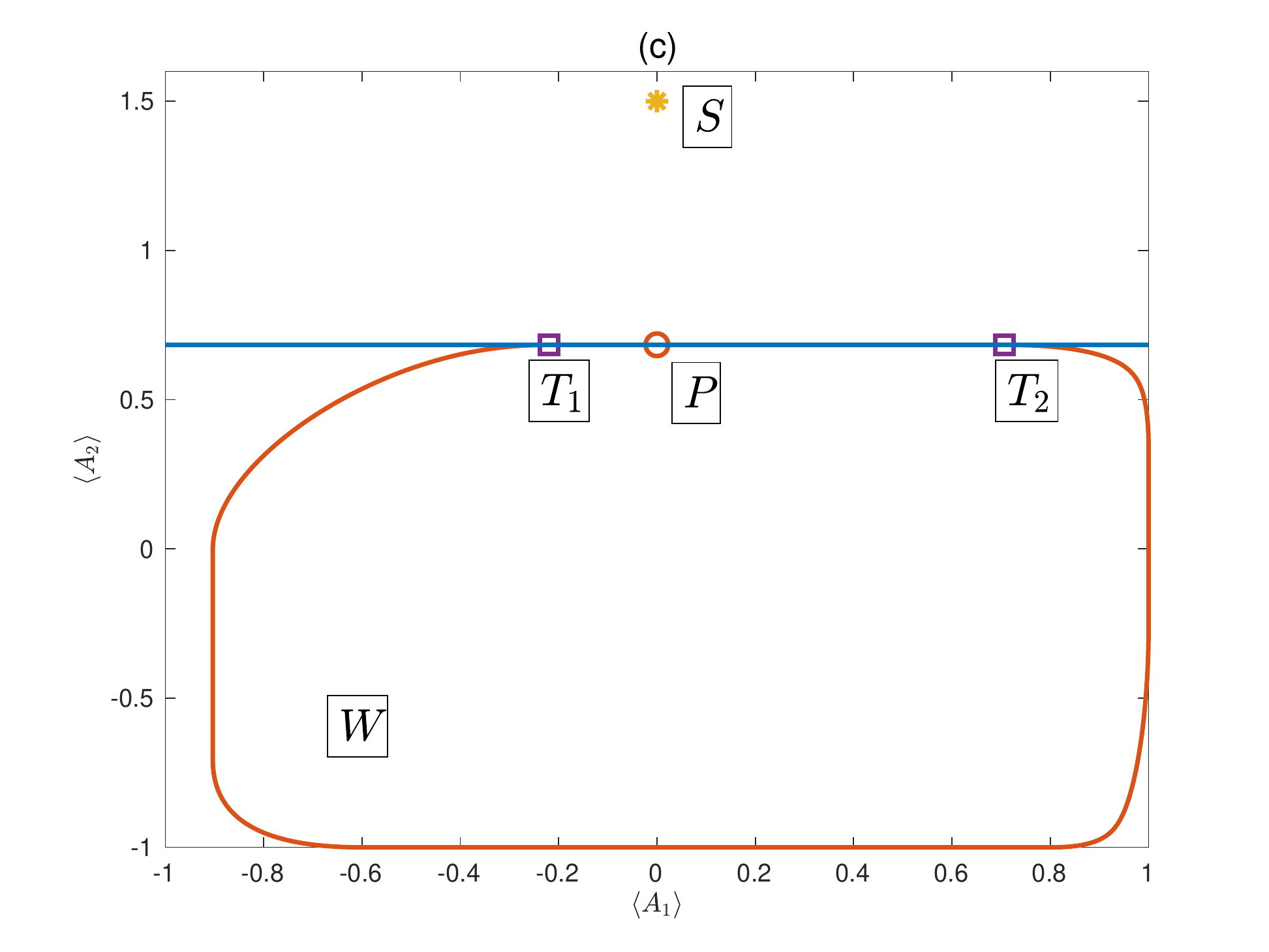}
\caption{\textbf{Result of the random operators} (a) The given point is inside the numerical range. (b) The given point is outside the numerical range without degeneracy. (c) The given point is outside the numerical range with degeneracy.}
\label{fig:2qubit}
\end{figure*}

As examples, we now consider the case of two qubits, i.e. $n=2$. And we choose $A_1=\sigma_x\otimes \sigma_x$ and $A_1=\sigma_z \otimes I_2$. For any state $\rho$, Let $\Tr(\rho A_1)$ be $x$-axis and $\Tr(\rho A_1)$ be $y$-axis. The boundary of $W(A_1,A_2)$ is a circle, whose center is $(0,0)$ and radius is $1$, as shown in \cref{fig:2qubit}.

Now, we have a state $\rho_1$ which satisfy $a_1=0.5$ and $a_2=0.5$. Obviously, the point $S(0.5,0.5)$ is inside $W(A_1,A_2)$, as shown in \cref{fig:2qubit}(a). The optimization will give a Hamiltonian 
\begin{equation}
H_1'=-2.4929A_1-2.4929A_2,
\label{eq:H1p}
\end{equation} corresponding the maximum entropy state
\begin{equation}
\rho_1'=
\begin{pmatrix}
0.3750    &     0     &    0  &  0.1250\\
         0   & 0.3750  &  0.1250     &    0 \\
         0  &  0.1250 &  0.1250     &    0\\
    0.1250    &     0   &      0  & 0.1250
\end{pmatrix},
\end{equation}
and one can readily check that $\Tr(\rho_1' A_1)=0.5$ and $\Tr(\rho_1' A_2)=0.5$, which is shown as point $P$ in \cref{fig:2qubit}(a). That is we found a maximum entropy state. You can see that $P$ and $S$ are almost the same points.
Also, the rank of  $\rho_1'$ is 4, which means $\rho_1'$ is a mixed state. The eigenvalues of $H_1'$ are $\{-0.8814,-0.8814,0.8814,0.8814\}$, and the energy gap between the ground state and the first excited states is $\thicksim 1.8$.

The state $\rho_2$, which satisfies $a_1=2$ and $a_2=2$, is obviously outside $W(A_1,A_2)$, as point $S$ shown in \cref{fig:2qubit}(b). 
The optimization will give a Hamiltonian $H_2'=-28.7177 A_1-28.7177 A_2$
and the corresponding maximum entropy state is 
\begin{equation}
\rho_2'=
\begin{pmatrix}
0.4268    &     0     &    0  &  0.1768\\
         0   & 0.4268  &  0.1768     &    0 \\
         0  &  0.1768 &  0.0732     &    0\\
    0.1768    &     0   &      0  & 0.0732
\end{pmatrix},
\end{equation}
and one can readily check that $\Tr(\rho_2' A_1)=0.7071$ and $\Tr(\rho_2' A_2)=0.7071$, which is shown as point $P$ in \cref{fig:2qubit}(a). 
One thing should be noticed here is that $\rank{\rho_2'}$ is 2, which means $\rho_2'$ is a mixed state of two degenerate ground states. These two ground states has the form:
\begin{equation}
|\psi_1\rangle=\begin{pmatrix}
0.9239\\0\\0\\0.3827
\end{pmatrix},
|\psi_2\rangle=\begin{pmatrix}
0\\0.9239\\0.3827\\0
\end{pmatrix}.
\end{equation}
and 
\begin{equation}
\rho_2'=0.5|\psi_1\rangle\langle\psi_1|+0.5|\psi_2\rangle\langle\psi_2|
\end{equation}
However, both $|\psi_1\rangle$ and $|\psi_2\rangle$ satisfy:
\begin{align}
\langle \psi_1 |A_1|\psi_1\rangle=\langle \psi_2 |A_1|\psi_2\rangle=0.7071,\\
\langle \psi_1 |A_2|\psi_1\rangle=\langle \psi_2 |A_2|\psi_2\rangle=0.7071.
\end{align}
Also, one can readily verify that $\Tr(\rho_2' A_1)=0.7071$ and $\Tr(\rho_2' A_1)=0.7071$. This is exactly the point on $W(A_1,A_2)$ which has the smallest Euclidean distance to the point $(2,2)$, as the point $P$ shown in \cref{fig:2qubit}(b).  Therefore, although $\rho_2'$ is degenerate, it is already an equally superpositioned state. Moreover, both the expectation of $|\psi_1\rangle$ and $|\psi_2\rangle$ corresponding to the same point, the point $P$, as shown in\cref{fig:2qubit}(b). The four eigenvalues of $H_2'$ is $\{-10.1532,-10.1532,10.1532,10.1532\}$ and the energy gap between the ground state and the first excited state is $\thicksim 20.3$, much greater than 1. Compared with the  energy gap of $H_1'$ in \cref{eq:H1p}, the energy gap of $H_2'$ is much greater, which guarantees that $\rho_2'$ is a mixed ground state.

With $H_2'$, we can draw a line through the point of $(0.7071,0.7071)$ with equation
\begin{equation}
\frac{x-0.7071}{28.7177}+\frac{y-0.7071}{28.7177}=0.
\label{eq:line}
\end{equation}
This line is a tangent line through point $(0.7071,0.7071)$. In \cref{fig:2qubit}(b), the function \cref{eq:line} represents the blue tangent line of the circle $W$.   Obviously, the points of $W(A_1,A_2)$ and the point $(2,2)$ lie at different sides of the line. Therefore, this line is a witness.

Now let us look at a case with flat boundary (corresponding to the degenerate ground states of some $H$). Suppose we have $3$ qubits and the subsystems are $\{1,2\}$ and $\{2,3\}$. Choose $A_1$ and $A_2$ random Hermitian operators acting on subsystem $\{1,2\}$ and $\{2,3\}$, respectively. For Hamiltonian
\begin{equation}
H=\sin\theta A_1 +\cos\theta A_2,
\end{equation}
the locality would guarantee that the ground state would be degenerate for some certain $\theta$, as shown in \cref{fig:2qubit}(c). Unlike the circle shown in \cref{fig:2qubit}(b), some parts of the boundary are flat, meaning that the ground state here is degenerate. Now, suppose we have a point $S(0,1.5)$, which is obviously outside $W$. The nearest point $P$, given by the maximum entropy state is two-degenerate: The rank of the density matrix $\rho_P$ corresponding $P$ is 2, which means there are two components in mixed state $\rho_P$:
\begin{equation}
\rho_P=p_1 |\psi_1\rangle\langle \psi_1|+p_2|\psi_2\rangle\langle \psi_2|,
\end{equation}
where $|\psi_1\rangle$ and $|\psi_2\rangle$ corresponds to point $T_1$ and $T_2$, respectively. Using the maximum entropy method again, we can find $T_1$ and $T_2$, as shown in  \cref{fig:2qubit}(c). Draw a line through  $T_1$ and $T_2$, this line is obviously a witness.

\section{The quantum marginal problem}\label{sec:qmp}

In this section, we apply our method to study a special case of quantum compatibility problem, namely the quantum marginal problem.
The quantum marginal problem asks whether there exists a global quantum state $\rho_N$, whose reduced density matrices $\tilde{\rho}_{S_i}=\Tr_{(S_i)^c}\rho_N$ on subsystem $S_i\subset \{1, 2, \cdots, N\}$\cite{klyachko2006quantum} coincide with given set$\{\rho_{S_i}\}$.
The related problem in fermionic (bosonic) systems is the so-called $N$-representability problem~\cite{coleman1963structure}.
It is worth mentioning that, the quantum marginal problem and the $N$-representative problem have been shown to be in the complexity class QMA, even for the relatively simple case in which all the given marginals are two-particle states~\cite{Liu06,LCV07,WMN10}.
The MaxExt then states that there exists a compatible $\rho_N$ if and only if there exists the $\bar{\rho}_N$ which has the maximum entropy among all the states that are compatible with the RDMs.

To be more precise, the quantum marginal problem could be defined as follows. Given a set of density matrices on several $2$-local systems $S_j$, such as $\{1, 2\}, \{2, 3\}, \{3, 4\}\cdots$, which are $\rho_{S_j}$ respectively.
Quantum marginal problem asks whether these density matrices are compatible, i.e. whether there exists a global state $\rho_N$, such that,
\begin{equation}
\Tr_{(S_j)^c}\rho_N=\rho_{S_j}, \qquad \forall S_j \subset \{1, 2, \cdots, N\}.
\label{eq:qmp}
\end{equation}
Denote $\Tr_{(S_j)^c}\rho_N$ by $\rho'_{S_j}$. If all $A_i$s in \cref{eq:ham} forms a complete basis of Hermitian operator on all $S_j$s, then \cref{eq:qmp} is established if and only if
\begin{equation}
\Tr(A_i \rho'_{S_j})=\Tr(A_i\rho_{S_j})=a_i,\, \forall A_is.
\end{equation}
Therefore, the problem is transformed into the compatibility problem:
for all $A_i$s, find a maximum entropy state
\begin{equation}
\rho'=\frac{e^{-H}}{\Tr e^{-H}},
\end{equation}
where
\begin{equation}
H'=\sum c_i A_i,
\end{equation}
so that
\begin{equation}
\Tr(A_i \rho'_{S_j})=\Tr(A_i\rho_{S_j})=a_i,\, \forall A_is.
\end{equation}
The maximum entropy method could be directly applied.

For quantum marginal problem, if the RDMs are not compatible, the final Hamiltonian $ H'$ is a witness. Let $E_g$ be the lowest eigenvalue of $H'$, the ground state energy. For any global state $\rho$, we have
\begin{equation}
\Tr (\rho H')\geq E_g.
\end{equation}
If the given RDMs $\{\rho_{S_i}\}$s are not compatible, then the Hamiltonian $H'$ obtained should satisfy
\begin{equation}
\sum_i H'_i \rho_{S_i}<E_g.
\end{equation}

We now consider some examples, starting from $n=3$.

\subsection{3-qubits}

It is known that for 3 qubit states,  nearly all pure states are uniquely determined by their 2-reduced density matrices among all states except the GHZ type states $a|000\rangle+b|111\rangle$ \cite{linden2002almost}.

First, let us look at the $W$-state:
\begin{equation}
|\psi_W\rangle=\frac{1}{\sqrt{3}}(|100\rangle+|010\rangle+|001\rangle
\end{equation}
By using the maximum entropy method, we will recover the result:
\begin{equation}
\rho'_{W}=\frac{e^{-H'_W}}{\Tr e^{-H'_W}}=
\begin{pmatrix}
0 &   0 &  0  &  0   & 0   & 0 &   0  & 0\\
    0  &  0.3333   & 0.3333  & 0 &   0.3333   &0   & 0 &  0\\
   0  &  0.3333   & 0.3333  & 0  &  0.3333  & 0  & 0  & 0\\
0 &   0 &  0  &  0   & 0   & 0 &   0  & 0\\
    0  &  0.3333  &  0.3333   & 0 &   0.3333   & 0   & 0  & 0\\
0 &   0 &  0  &  0   & 0   & 0 &   0  & 0\\
0 &   0 &  0  &  0   & 0   & 0 &   0  & 0\\
0 &   0 &  0  &  0   & 0   & 0 &   0  & 0
   \end{pmatrix},
\end{equation}
which is the density matrix of $W$-state. The lowest two eigenvalues of the corresponding Hamiltonian $H'_W$ are $-28.8105$ and $-18.3106$, which guarantees that the rank of the density matrix $\rho'_{W}$ is 1, hence a pure ground state.

Now, suppose we have the GHZ-state:
\begin{equation}
|\psi_{\text{GHZ}}\rangle=\frac{1}{\sqrt{2}}(|000\rangle+|111\rangle).
\end{equation}
This time, the maximum entropy state will give a density matrix
\begin{equation}
\rho'_{\text{GHZ}}=0.5(|000\rangle\langle 000|+|111\rangle\langle 111|).
\end{equation}
It can be easily verified that $\rho'_{\text{GHZ}}$ and $|\psi_{\text{GHZ}}\rangle\langle \psi_{\text{GHZ}}|$ have the same 2-body reduced density matrix.  Obviously, the populations of the two components - $|000\rangle$ and $|111\rangle$ - in state $\rho'_{\text{GHZ}}$ are both 0.5. For a maximum entropy state, or a thermal state, the only possibility that two components have the same population is that these two components are degenerate. Let the corresponding Hamiltonian of $\rho'_{\text{GHZ}}$ be $H'_{\text{GHZ}}$, we have
\begin{align}
H'_{\text{GHZ}}|000\rangle&=-32.2662 |000\rangle,\\
H'_{\text{GHZ}}|111\rangle&=-32.2662 |111\rangle.
\end{align}
Both $|000\rangle$ and $|111\rangle$ have the same eigen-energy, which means they are both ground state of the Hamiltonian $H'_{\text{GHZ}}$.

\subsection{4-qubit chain}

Now suppose we have a system with 4 qubits labelled as 1,2,3,4, and  the subsystems are
\begin{align}
S_1&=\{1,2\};\\
S_2&=\{2,3\};\\
S_3&=\{3,4\},
\end{align}
Now the problem is that given the reduced density matrix $\rho_{12},\rho_{23}$ and $\rho_{34}$, determine whether there exist a global density matrix, of which $\rho_{12},\rho_{23}$ and $\rho_{34}$ are the reduced density matrices.

Suppose we have a bell state
\begin{equation}
|\psi_{B}\rangle=\frac{1}{\sqrt{2}}(|00\rangle+|11\rangle),
\end{equation}
with the density matrix
\begin{equation}
\rho_{B}=|\psi_B\rangle\langle\psi_B|.
\end{equation}
Suppose $\rho_{12}=\rho_{23}=\rho_{34}=\rho_{B}$. Because of entanglement monogamy, we know that there does NOT exist a global $\rho$, of which the reduced density matrices are $\rho_{12},\rho_{23}$ and $\rho_{34}$.

Our maximum entropy method gives a density matrix $\rho'=e^{-H'}/\Tr(e^{-H'})$, of which the rank is 1. This means $\rho'$ is a pure state. It can be verified that $\rho'$ is the eigenstate with the lowest eigen-energy of $H'$ - the ground state. The lowest energy of $H'$  is $-81.1733$. Therefore, for any global state $\rho$, we have
\begin{equation}
\Tr(\rho H')\geq -81.1733.
\end{equation}
Also, we know that $H'$ is a local Hamiltonian:
\begin{equation}
H'=H'^{g}_{12}+H'^{g}_{23}+H'^{g}_{34},
\end{equation}
where the superscript $g$ means the local interaction Hamiltonian in global form. Now, we can verify that
\begin{equation}
\Tr(\rho_{12}H'_{12})+\Tr(\rho_{23}H'_{23})+\Tr(\rho_{34}H'_{34})=-121.0888,
\end{equation}
which is smaller than the ground state energy of $H'$. Therefore, $H'$ is a witness.

The above case gives a very example that $H'$ is not degenerate. However, if you randomly choose $\rho_{12}, \rho_{23}$ and $\rho_{34}$, the solution is mostly likely to be degenerate. For example, we randomly generate the three reduced density matrices as follows
\begin{widetext}
\begin{align}
\rho_{12} &=\begin{pmatrix}
0.2408 &0.1717 - 0.1312i & -0.1304 - 0.0459i  & 0.0306 - 0.0962i\\
   0.1717 + 0.1312i &   0.3359 &  -0.0536 - 0.1599i  & 0.1028 - 0.0836i\\
  -0.1304 + 0.0459i  &-0.0536 + 0.1599i  & 0.3336&  0.0695 + 0.1169i \\
   0.0306 + 0.0962i  & 0.1028 + 0.0836i   &0.0695 - 0.1169i  & 0.0896 + 0i
\end{pmatrix};\\
\rho_{23}& =\begin{pmatrix}
 0.2534 & -0.0662 + 0.0248i &  0.1068 - 0.0440i  & 0.0746 - 0.1027i\\
  -0.0662 - 0.0248i  & 0.2168  & 0.0307 - 0.0383i  &-0.1085 + 0.2051i\\
   0.1068 + 0.0440i &  0.0307 + 0.0383i  & 0.2114 &   0.0309 - 0.0164i\\
   0.0746 + 0.1027i & -0.1085 - 0.2051i &  0.0309 + 0.0164i &  0.3183
\end{pmatrix};\\
\rho_{34} &=\begin{pmatrix}
0.1175&-0.0402 - 0.0453i & -0.0265 + 0.0143i  & 0.0747 + 0.0688i\\
  -0.0402 + 0.0453i &  0.5037 & -0.1095 + 0.0187i & -0.0620 - 0.1296i\\
  -0.0265 - 0.0143i & -0.1095 - 0.0187i &  0.0883 &  -0.0247 - 0.0074i\\
   0.0747 - 0.0688i & -0.0620 + 0.1296i & -0.0247 + 0.0074i  & 0.2905
   \end{pmatrix}.
\end{align}
\end{widetext}
The maximum entropy method will give a $\rho'$, of which the rank is 8, meaning that $\rho$ is a mixture of 8 pure states. As mentioned earlier, for states outside $\mathcal{B}$ shown in \cref{fig:geoint}, the nearest points is on the boundary, which is composed of ground states. Therefore, the 8 pure states are all ground states of some Hamiltonian $H'$, and $H'$ is degenerate: all the 8 pure states are ground states and have the same energy.

To find the Hamiltonian $H'$, we generate a new density matrix:
\begin{equation}
\rho_{g}=\sum_i \frac{1}{n}|\psi_i\rangle\langle \psi_i|,
\end{equation}
where $n$ is the degeneracy and $|\psi_i\rangle$ is the $i$-th ground states in $\rho''$. By using the maximum entropy method, we can find the Hamiltonian $H''$, of which
\begin{equation}
\rho''=\frac{e^{-H''}}{\Tr e^{-H''}}.
\end{equation}
In our example, the lowest eigen-energy of $H''$ is around -7.2: because of numerical precision, there are slightly difference between these energies. The first exited states have the energy of 1.7, which makes the gap between the ground state be about 8.9. This gap makes the state $\rho''$ a ground state.

Similarly, we could calculate the energy of $\{\rho_{12},\rho_{23},\rho_{34}\}$ based on $H"$:
\begin{equation}
\Tr(\rho_{12}H''_{12}+\Tr(\rho_{23}H''_{23})+\Tr(\rho_{34}H''_{34})=-7.5,
\end{equation}
which is smaller than the ground state energy of $H''$. Therefore, $H''$ is a witness.

\section{The hybrid quantum-classical algorithm}\label{sec:var}
In our algorithm, we need to repeatedly calculate density matrices of the form
\begin{equation}
\rho=\frac{e^{-H}}{\Tr(e^{-H})}.
\label{eq:target}
\end{equation}
Calculation of the exponential of a matrix using a classical computer is a difficult task, because it is not scalable. To address this issue, we will turn to some hybrid quantum-classical algorithm. Recently,  there have been some proposed hybrid quantum-classical algorithms~\cite{zeng2021VQHD, verdon2019quantum, wu2019variational, wang2021variational, Chowdhury2020VQTS, zoufal2021variational} to prepare the thermal state $\rho_{\beta}=\frac{e^{-\beta H}}{\Tr(e^{-\beta H})}$.  All of these can be used to help scale our method with the existence of near-term quantum computers.

As an example, we have used the method in Ref.~\cite{zeng2021VQHD} to calculate the thermal state density matrix, which can take advantage of the small correlation length of the system in small $\beta$.  When  setting $\beta=1$,  we can obtain the density matrix in Eq.~\ref{eq:target}.
To prepare the thermal state $\rho_{\beta}$, one can apply an imaginary time evolution with a thermofield double state.  We first prepare initial state $\ket{\phi_0}  = \frac{1}{\sqrt{2^n}} \sum_{i=1}^{2^n}\ket{i}\ket{i}$ with  $2n$ qubits. The imaginary time evolution $e^{-\beta H/2}$ on $\ket{\phi_0}$ returns the state $\ket{\text{TFD}(\beta)} =  \sqrt{\frac{2^n}{\Tr(e^{-\beta H})}} e^{-\beta H/2} \ket{\phi_0}$ and $\rho_{\beta}$ will be then obtained on the first $n$ qubits by tracing out the last $n$ qubits, $\rho_{\beta}=\Tr_{n+1,\ldots ,2n}\ket{\text{TFD}(\beta)}\bra{\text{TFD}(\beta)}$.

To implement $e^{-\beta H/2}$ on a quantum computer, where $H=\sum_m h[m]$ acts nontrivially only on at most $K$ neighbouring qubits, one could use a quantum imaginary time evolution (QITE) algorithm, which is NISQ friendly~\cite{motta2020determining, cao2022quantum}. The basic idea of QITE is to approximate the $K$-local non-unitary transformation with a $D$-local unitary transformation in each step after the Trotter decomposition,
\begin{eqnarray}
|{\psi}' \rangle = \frac{1}{\sqrt{c}} e^{-\Delta\tau h[m]} |\psi \rangle \approx e^{-i\Delta\tau A[m]} |\psi \rangle,
\label{basicQite}
\end{eqnarray}
where $c = \langle \psi|e^{-2\Delta\tau h[m]} |\psi \rangle$ is the normalization factor.
Each $h[m]$ is a $K$-local operator and $A[m]$ is Hermitian, which can be expanded in terms of Pauli basis on $D$ qubits,
\begin{eqnarray}
A[m] = \sum_{i_1 i_2 ... i_D} a[m]_{i_1 i_2 ... i_D} \sigma_{i_1} \sigma_{i_1} ... \sigma_{i_D} = \sum_{I} a[m]_{I} \sigma_{I},
\end{eqnarray}
where $a[m]_{I}$ is the coefficient of combining Pauli operator $\sigma_{I}$ and the index $I$ is a combination of qubit indexes $\{i_1,i_2, ... ,i_D\}$ corresponding to nontrivial Pauli operators. To find coefficients $a[m]_{I}$ of $A[m]$,  we minimize the square norm of the difference between the two quantum states
\begin{eqnarray}
\parallel |{\psi}' \rangle -(1-i\Delta\tau A[m]) |{\psi} \rangle \parallel^2,
\end{eqnarray}
where the $(1-i\Delta\tau A[m]) |{\psi} \rangle$ is the first two terms of the taylor series of $e^{-i\Delta\tau A[m]} |\psi \rangle$.  The solution of the minimization is subject to the linear equation,
\begin{eqnarray}
\label{linear-eq}
(\boldsymbol{S} + \boldsymbol{S}^T) \boldsymbol{a}[m] = -\boldsymbol{b},
\end{eqnarray}
where the matrix $\boldsymbol{S}$ and vector $\boldsymbol{b}$ can be obtained by $D$ local measurements on the quantum state $|{\psi} \rangle$,
\begin{eqnarray}
\label{eq:Sb-matrix}
S_{IJ} &=& \langle \psi| \sigma_{I}^{\dagger} \sigma_{J}  |\psi \rangle , \nonumber \\
b_{I}  &=& -  2\text{Im}\left[\frac{1}{\sqrt{c}}\langle \psi| \sigma_{I}^{\dagger} h[m] | \psi \rangle \right],
\end{eqnarray}
where Im[] denotes the imaginary part of the variable inside.
We can obtain the $\boldsymbol{a}[m]$ by solving the linear equation on a classical computer, and then construct a quantum circuit to implement the unitary transformation $e^{-i\Delta\tau A[m]} |\psi \rangle$ on an NISQ quantum devices.

The locality of $D$ qubits local unitary transformation is important. The unitary local operator $A[m]$ should at least act nontrivially on the $(i, i+n)$ qubit pair if its counterpart $h[m]$ act nontrivially on $i$-th qubit~\cite{zeng2021VQHD}.
Given a $K$-local Hamiltonian, the QITE algorithm is capable of capturing the correlation of the original Hamiltonian only if $D\geq 2K$.
In our case  it is a $2$-local Hamiltonian, so we set $D=4$.
The trotter step size $\Delta\tau$ can also affect the accuracy of the thermal state density matrix, therefore we here choose a small value $\Delta\tau=0.05$.
Then the number of Trotter steps is $N=\frac{\beta/2}{\Delta\tau}=10$.
As described in Ref.~\cite{zeng2021VQHD} the  time complexity of the QITE and the depth of the circuit construct by the QITE algorithm depend on $N \times \mathcal{O}(e^{D})$.
The density matrix in Eq.~\ref{eq:target} thus can be calculated efficiently on NISQ devices.

In previous section we showed the results in the 3-qubit case, where for GHZ state, the variation method returns the density matrix
\begin{equation}
\rho_{\text{GHZ}}=0.5(|000\rangle\langle 000|+|111\rangle\langle 111|).
\end{equation}
and for W state, it returns exactly the target density matrix, which means that our variational method works as desired.

\section{Conclusion}\label{sec:con}

In this work, we applied the maximum entropy method to solve the quantum state compatibility problem.
Compared to the traditional SDP method, there are several advantages of the method.
At first, only a small number of parameters (the same as the number of measurement operators) are required to parametrize to obtain the desired density matrix.
Secondly, the hybrid quantum-classical algorithms replace some subroutines with help of quantum devices may give further advantages with the incoming NISQ era.
Thirdly, in case the solution does not exist, our method will further produce a witness which is the supporting hyperplane of the compatible set; making use of these "byproducts" might be able to provide us some insight in further understanding about the global features of the compatible sets in future studies.

\section{Acknowledgement}

S-Y.Hou is supported by National Natural Science Foundation of China under Grant No. 12105195. YN-Li is supported by National Natural Science Foundation of China under Grant No. 12005295.  Z-P. Wu, C.-F. Cao, B.Zeng. are supported by GRF16305121.

\bibliographystyle{apsrev4-1}
\bibliography{references}

\end{document}